\documentclass[reprint,superscriptaddress,amsmath,amssymb,aps,prl,floatfix]{revtex4-2}

\usepackage[utf8]{inputenc}
\usepackage{graphicx}
\usepackage{bm}
\usepackage{siunitx}
\usepackage[left=2.5cm, right=2.5cm, top=2.5cm]{geometry}
\usepackage{amsmath,amssymb}
\usepackage{color}
\usepackage{comment}
\usepackage{xcolor}
\usepackage{float}

\definecolor{darkblue}{rgb}{0.,0.,0.6}
\usepackage[colorlinks=true,allcolors=darkblue]{hyperref}%
\usepackage{booktabs} 

\usepackage[mode=buildnew]{standalone}
\usepackage{tikz}

\newcommand{\bra}[1]{\langle #1|}
\newcommand{\ket}[1]{|#1\rangle}

\newcommand{\Yb}{Yb$^+$\,}
\newcommand{\Ybf}{$^{171}$Yb$^+$\,}
\newcommand{\Ybc}{$^{172}$Yb$^+$\,}

\newcommand{\Bs}{{^{1}{[5/2]}_{5/2}}}
\newcommand{\Bso}{{^{1}[5/2]^{o}_{5/2}}}

\newcommand{\Dfive}{{^{2}{D}_{5/2}}}
\newcommand{\Dthree}{{^{2}{D}_{3/2}}}
\newcommand{\Shalf}{{^{2}{S}_{1/2}}}
\newcommand{\Phalf}{{^{2}{P}_{1/2}}}
\newcommand{\Fseven}{{^{2}{F}_{7/2}}}
\newcommand{\ttau}{$\tau=37.9(9) \,\mu$s\,}

\begin{document}
\preprint{APS/123-QED}

\title{Precision Measurement of Lifetime and Branching Ratios\\ of the $4f^{13}5d6s\,^1[5/2]_{5/2}$ state in Yb$^+$ ions}

\author{Midhuna Duraisamy Suganthi}
\email{md76@rice.edu}
\affiliation{Department of Physics and Astronomy and Smalley-Curl Institute, Rice University, Houston, TX 77005, USA}
\affiliation{Applied Physics Graduate Program, Smalley-Curl Institute, Rice University, Houston, TX 77005, USA }
\author{Visal So}
\affiliation{Department of Physics and Astronomy and Smalley-Curl Institute, Rice University, Houston, TX 77005, USA}
\author{Abhishek Menon}
\affiliation{Department of Physics and Astronomy and Smalley-Curl Institute, Rice University, Houston, TX 77005, USA}
\author{George Tomaras}
\affiliation{Department of Physics and Astronomy and Smalley-Curl Institute, Rice University, Houston, TX 77005, USA}
\affiliation{Applied Physics Graduate Program, Smalley-Curl Institute, Rice University, Houston, TX 77005, USA }
\author{Roman Zhuravel}
\affiliation{Department of Physics and Astronomy and Smalley-Curl Institute, Rice University, Houston, TX 77005, USA}
\author{Guido Pagano}
\email{pagano@rice.edu}
\affiliation{Department of Physics and Astronomy and Smalley-Curl Institute, Rice University, Houston, TX 77005, USA}

\begin{abstract}
We report spectroscopic and time-resolved experimental observations to characterize the $[{\rm Xe}]4f^{13}(^2F^{o}_{5/2}){5d6s(}{^1\!D})\Bso$ state in \Ybc ions. We access this state from the metastable $4f^{14}5d (^2D_{3/2,5/2})$ manifold and observe an unexpectedly long lifetime of \ttau that allows visible Rabi oscillations and resolved-sideband spectroscopy. Using a combination of coherent population dynamics, high-fidelity detection and heralded state preparation, and optical pumping methods, we measure the branching ratios to the $^{2}D_{3/2}$, $^2D_{5/2}$, $^2S_{1/2}$ states to be $0.359(2)$, 0.639(2), $0.0023(16)$, respectively. The branching ratio to the $4f^{13}6s^{2}({^2F}_{7/2})$ is compatible with zero within our experimental resolution. We also report measurements of its Landé g-factor and the branching ratio of the $\Dfive$ to $\Shalf$ decay in \Ybc to be 0.188(3), improving its relative uncertainty by an order of magnitude. Our measurements pave the way to a better understanding of the atomic structure of \Yb ions, which still lacks accurate numerical descriptions, and the use of high-lying excited states for partial detection and qubit manipulation in the \emph{omg} architecture.
\end{abstract}

\maketitle

Trapped ions are a pristine quantum platform used for diverse applications, including quantum information processing \cite{Bruzewicz2019}, quantum simulation \cite{monroe2021programmable}, and quantum metrology \cite{Ludlow2015}. In particular, \Yb ions are a workhorse for these applications: they feature long-lived quantum memory \cite{Wang2021single}, high-fidelity single- and two-qubit gates \cite{Moses2023}, and are routinely used for the quantum simulation of spin \cite{tan2021domain,de2024observationstringbreakingdynamicsquantum, joshi2022observing} and spin-boson models \cite{sun2024quantumsimulation, so2024trapped, so2025}. \Yb ions also provide multiple clock transitions \cite{PhysRevLett.108.090801}, making them one of the most attractive ions to develop optical frequency standards \cite{STUHLER2021100264}. Additionally, they feature a large number of stable isotopes, making them a promising candidate for new boson searches via non-linearities in King's plots \cite{Berengut2020, Hur2022, Door2025}.

\begin{figure}[t!]
\centering
\includegraphics[width=0.48\textwidth]{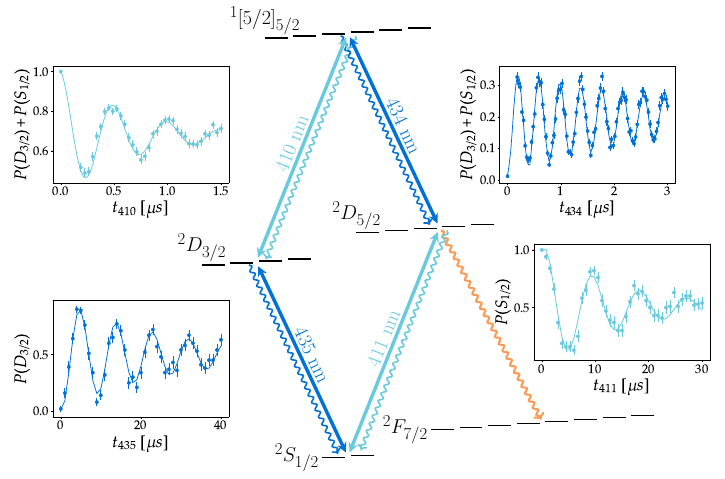}
\vspace{-0.7cm}
\caption{ {\bf Level scheme}: Rabi oscillations observed on $\Shalf\!\leftrightarrow\!\Dthree(\Dfive)$ and $\Dthree(\Dfive)\!\leftrightarrow\!\Bs$ transitions. The maximum population inversion of the $\Dthree(\Dfive)\!\!\rightarrow\!\!\Bs$ transitions are limited by the total branching ratio to the $\Dthree$ and $\Shalf$ states, $b_{D_{3/2}}+b_{S_{1/2}}=0.361(2)$.  
The solid lines are the OBE fits with the Rabi frequencies as free parameters, resulting in $\Omega/2\pi = {(110.3,112.3,2609.5,2013.8)}$ kHz for the four (435 nm, 411 nm, 434 nm, 410 nm) transitions (see Supplemental Material for the relevant pulse sequences). Each data point is the average of 500 experimental runs, and the error bars are $\pm$1 standard deviation from the mean. 
}
\vspace{-0.5cm}
\label{fig:Yblevel}
\end{figure}

Despite its prominence in many applications, the atomic levels of singly ionized Yb atoms still cannot be accurately predicted by existing numerical methods \cite{Porsev2012, 
Kahl2019}
due to their extremely complex electronic structure that involves 13 or 14 electrons occupying the inner-shell $4f$ orbital. Since low-lying core-excited configurations, such as $4f^{13}6s^2$ and $4f^{13}5d6s$, strongly mix with valence configurations, core-valence correlations have to be accurately taken into account to predict the matrix elements and energy levels \cite{Migdalek2000, Safranova2009, Porsev2012, Migdalek_2012, Radziute2021}.
For example, the theoretical prediction of the lifetime of the $4f^{13}6s^2(F_{7/2})$ state \cite{Dzuba2016, Guo2020} is still not in agreement with the latest measured experimental value \cite{Lange2021lifetime}. In the case of high-lying excited states, predictions of the matrix elements are even more challenging because an untreatable number of electron configurations, including both valence and core-excited ones, strongly mix and need to be considered \cite{Biemont1998}.

One of these high-lying atomic levels, $4f^{13}(^2F^o_{5/2})5d6s({^1D})\,\Bso$ (in the following, abbreviated as $\Bs$), has recently attracted interest \cite{Schacht2015Yb} as it could provide a convenient way to dissipatively initialize the metastable state qubit in the $omg$ architecture with \Ybf \cite{Allcock2021omg}, potentially faster than relying on the natural decay from $\Dfive$ state to $\Fseven$ state \cite{Conrad2021, Edmunds2021}.
The $\Bs$ state forms an almost perfect ``rhomboid'' of frequencies with the $^2S_{1/2}$, $^2D_{3/2}$, and $^2D_{5/2}$ states as all the four transitions connecting them lie in the same spectral region (410, 411, 434, and 435 nm) and, therefore, can be excited by tuning the same lasers used for the E2 $\Shalf\!\!\leftrightarrow{\!\!\Dthree,\Dfive}$ transitions (see Fig.~\ref{fig:Yblevel}). In Ref.~\cite{fourmetastable410}, the absolute frequencies, isotope shifts, and hyperfine splittings of the $\Bs$ state have recently been measured by exciting the $\Dthree\rightarrow{\Bs}$ transition at 410 nm. However, no detectable branching ratio to the $\Fseven$ state has been reported.

In this work, we report the direct measurement of the lifetime and branching ratios of the $\Bs$ state by using a combination of heralded state preparation, optical pumping, and coherent drive of the  
$\Dfive\leftrightarrow{\Bs}$ and $\Dthree\leftrightarrow {\Bs}$ transitions at 434 nm and 410 nm, respectively (see Fig.~\ref{fig:Yblevel}). The $\Bs$ state is thought to have a 0.12 mixing with the $4f^{13} (^2F_{7/2}) 5d^2 (F_3)$ configuration \cite{Ralchenko2005,3974EL}. Therefore, the naive expectation is that the $^2D_{3/2,5/2}\leftrightarrow{^1[5/2]_{5/2}}$ transitions are dipole-allowed since they also have opposite parity and $\Delta J=0,1$. However, we measure an unexpectedly long lifetime
(\ttau\!), which allows the observation of distinct Rabi oscillations on both $^2D_{3/2,5/2}\leftrightarrow \Bs$ transitions, when driving them with a sufficiently high-power, narrow-linewidth laser, locked to an ultralow-expansion (ULE) cavity. We further quantify the branching ratios to the ${\Dthree,\Dfive,\Shalf,\Fseven}$ states and also perform a measurement of the Landé factor $g_J$ of the $\Bs$ state.

\begin{figure*}[t!]
    \centering
    \includegraphics[width=\textwidth]{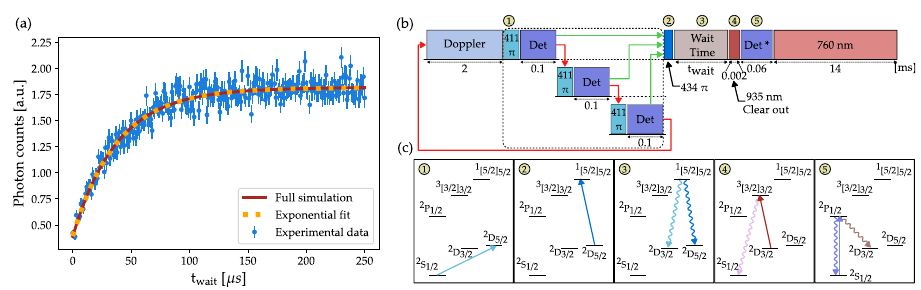}
    \vspace{-0.7cm}
    \caption{
    {\bf(a)} Lifetime measurement: Observation of the decay from $\Bs$ state to $\Dthree$ and $\Shalf$ states as a function of $t_{\rm wait}$. The solid line is the solution of the OBEs using QuTiP. A scaling parameter and an offset are added to the simulation result to fit the unthresholded photon counts of the experimental scan. Each data point is the average of 2000 experimental repetitions, and the error bars are $\pm$1 standard deviation from the mean.
    {\bf(b,c)} Pulse sequence consisting of the heralded 411 nm state-preparation steps, 434 nm $\pi$ pulse (0.33 $\mu$s), variable wait time $t_{\rm wait}$, 935 nm clearout pulse (2 $\mu$s), and detection without 935 nm step (60\,$\mu$s, labeled as $\rm {Det}^*$).
    }
    \label{fig:lifetime}
    \vspace{-0.5cm}
\end{figure*}

\emph{Lifetime measurement} - The experiments are performed on a single \Ybc ion confined in a Paul trap with radial trap frequencies $(\omega_x,\omega_y)/2\pi=(3.31,3.86)$ MHz and axial trap frequency $\omega_z/2\pi=0.93$ MHz (see Ref.~\cite{so2024trapped} for more details). The ion is Doppler-cooled using the near-cycling transition $\Shalf\leftrightarrow\Phalf$ at 370 nm, together with a 935 nm repumper addressing the $\Dthree\leftrightarrow {^1[3/2]_{3/2}}$ transition (see Fig.~\ref{fig:lifetime} and Ref.~\cite{Olmschenk2007}). Doppler cooling leads to a thermal population of $\bar{n}\sim 6.5$ in the radial and $\bar{n}\sim 22$ in the axial directions, which causes the decay of the oscillations in Fig.~\ref{fig:Yblevel}.

Two laser beams at 411 nm (435 nm) and 434 nm (410 nm) are used to resonantly address the $\Shalf\leftrightarrow\Dfive\,(\Dthree)$ and the $\Dfive\,(\Dthree)\leftrightarrow\Bs$ transitions, respectively. Both lasers are PDH-locked \cite{Black2001} to the ULE cavity with a finesse of 9500. Since the $\Dfive$ state has a 81.2\% branching ratio to the $\Fseven$ state \cite{Taylor1997}, a 760 nm repumper to the $4f^{13}(^2F^o_{7/2})5d6s({^1\!D})\,^1[3/2]^o_{3/2}$ state is used at the end of every experiment to re-initialize the population in the $\Shalf$ ground state \cite{Sugiyama_1999, Edmunds2021}. The population in the $\Shalf$ and $\Dthree$ states is measured by collecting the ion fluorescence with a 0.6 NA objective under 370 nm and 935 nm illumination for 100 $\mu$s. In this case, the Poissonian distributions of the detected photons for bright ($\Shalf$ and $\Dthree$) and dark ($\Dfive, \Fseven, \Bs$) states are well separated, allowing the use of threshold discrimination. Conversely, the $\Shalf$ population alone can be detected at the end of the experiment by illuminating the ion only with the 370 nm beam for 60 $\mu$s, which results in the collection of the photons scattered during the optical pumping to the $\Dthree$ state.

The pulse sequence used to measure the decay rate of $\Bs$ consists of the following steps (see Fig. \ref{fig:lifetime}c): 
\emph{(i)} The ion is prepared in the $\ket{m_j=-1/2}$ Zeeman sublevel of $\Dfive$ with a 411 nm $\pi$ pulse. To significantly increase the state-preparation fidelity, we use a heralding scheme by applying a detection step with 370 nm and 935 nm, measuring the population in $\Shalf$ and $\Dthree$ states after the 411 nm $\pi$ pulse.
If the 370 nm photon counts are below the threshold, the $\Dfive$ state preparation is considered to be successful, and we move on to the next experimental pulse. Conversely, if the 370 nm detection photon counts exceed the threshold, the heralding scheme is iterated until the state preparation is successful. This procedure results in a dark fidelity of $99.7(1)\%$, corresponding to a 98.7(1)\% fidelity in the $\Dfive$ state, which is limited primarily by the decay from the $\Dfive$ state during the 0.1 ms detection time.
\emph{(ii)} 
A 434 nm $\pi$ pulse ($0.33\,\mu$s) is applied to populate the $\ket{\Bs, m_j=-3/2}$ Zeeman sublevel. 
\emph{(iii)} Then, a wait time is scanned to allow the decay of the population from the $\Bs$ state. 
\emph{(iv)}
We then measure the $\Dthree$ population by first transferring it to the $\Shalf$ state with a 935 nm clearout pulse (2 $\mu$s). This is much faster than the overall $\Bs\!\!\rightarrow\!\{\Dfive,\Dthree\}$ decay timescale and thus sufficient to completely transfer the $\Dthree$ population to the $\Shalf$ state. \emph{(v)} Finally, we detect only the $\Shalf$ population using the 370 nm light for $60\, \mu$s. This procedure is adopted because the 100 $\mu$s detection of both $\Shalf$ and $\Dthree$ populations is longer than the decay lifetime $\tau$ and, therefore, would affect the decay measurement. A fast population transfer from $\Dthree$ to $\Shalf$ has the advantage of allowing a direct measurement in real time of the decay probability, but it has the drawback that the number of photon counts acquired during detection is low, hindering the application of thresholding.

The raw photon counts as a function of wait time are reported in Fig.~\ref{fig:lifetime}. We fit the data with both an exponential function and with a QuTiP \cite{johansson2013qutip} numerical integration of the optical Bloch equations (OBEs) (see Supplemental Material) to take into account the finite 935 nm clearout time (2 $\mu$s) and the $\Dfive$ and $\Dthree$ decays to the $\Shalf$ and $\Fseven$ states. The two fitting methods agree within 1\%, showing that the 935 nm clearout time and the decays from the $\Dfive$ and $\Dthree$ metastable states are much faster and slower than the decay time, respectively. The fit of the data to the full simulation results in the $\Bs$ state lifetime of \ttau, which is the result of five independent measurements with the error bar being the weighted standard deviation (see Appendix).

\begin{figure*}[t!]
\centering
\includegraphics[width=\textwidth]{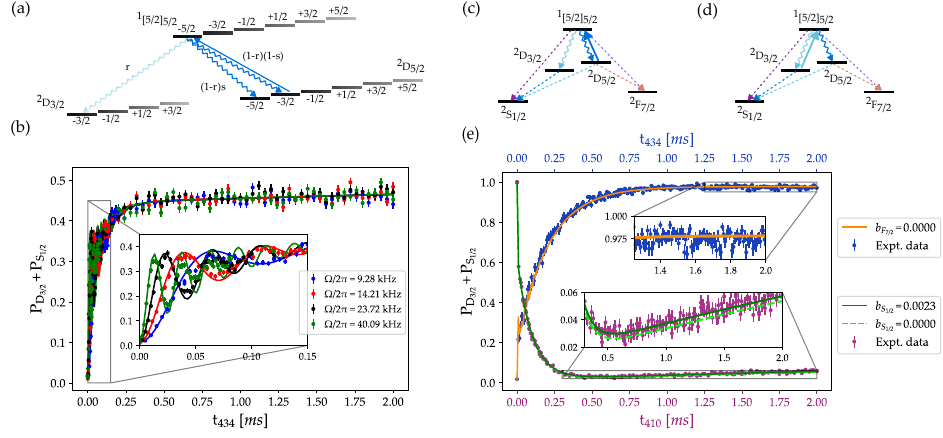}
\vspace{-0.8cm}
\caption{Branching ratio measurements: 
{\bf(a)} Level scheme highlighting the Zeeman sublevels involved in the experiment. $r=b_{D_{3/2}}$ is the branching ratio to the $\Dthree$ state and $s=b_{D_{5/2}}\cdot cb$, where $cb$ is the Clebsch-Gordan coefficient (see text).
{\bf (b)} Low-power evolutions at $B=4.839(1)$ Gauss as a function of $t_{434}$ for four different laser powers. The independently calibrated Rabi frequencies ($\Omega_{434}/2\pi = 9.28, 14.21, 23.72, 40.09$ kHz) are in the inset.
{\bf (c,d)} Level scheme for low magnetic field ($B=0.448(2)$ Gauss) with high-power 434 nm (c) and 410 nm (d) drives, corresponding to blue and purple points in (e), respectively. 
{\bf (e)} High-power and low-magnetic-field evolutions as a function of $t_{434}$ (blue points) and $t_{410}$ (purple points) with $\Omega_{434}/2\pi\approx7800$ kHz, and $\Omega_{410}/2\pi\approx2800$ kHz. The solid lines are fits to the OBEs with the only fit parameter being $b_{S_{1/2}}$ and $b_{F_{7/2}}$, respectively. The dashed green line is the result of the simulation assuming $b_{S_{1/2}}=0$.
In plots, each point is the average of 500 (blue) and 2000 (purple) repetitions, with the error bars being $\pm$1 standard deviation from the mean. 
}
\label{fig:Branching}
\end{figure*}

In order to make the signal as stable as possible, we set the Rabi frequency of the 434 nm pulse $\Omega_{434}/2\pi= 1.519$ MHz: on the one hand, this allows us to address a single Zeeman state as $\Omega_{434}/2\pi<\Delta f_{\rm Zeeman}=6.806$ MHz; on the other hand, it makes the pulse sequence largely insensitive to the ULE cavity frequency drifts (0.175 Hz/s) and the magnetic field fluctuations (see Appendix). However, the fluctuations in the raw detection counts and the imperfect $\pi$ pulses due to 434 nm laser intensity fluctuations cause the statistical error bar to underestimate the error in each data point. This results in an average reduced $\chi^2/(N-1)=4.37$ over the five measurements. We take this into account by rescaling the statistical error bar of each point in Fig.~\ref{fig:Branching} to set $\chi^2/(N-1)=1$, as suggested by PDG \cite{PhysRevD.110.030001}.

\emph{Branching ratio measurements} - Precision measurements of the branching ratios have been done in trapped ions with multi-level systems by observing the scattering fluorescence given by the population being pumped to a dark state or by directly measuring the fraction of population in the pumped dark state \cite{Gerritsma2008, Ramm2013, Zhang2016, Arnold2019}. 
Collecting the emitted 410 nm or 434 nm photons would result in very long experimental runs because of the very low scattering rate ($\gamma_{[5/2]}/2\pi=1/2\pi\tau=4.2(1)$ kHz) and the fact that we can collect only $\sim10$\% of the emitted photons. 
For our measurement, we prepare the population in $\ket{\Dfive,m_j=-3/2}$ with the heralding protocol and drive the $\ket{\Dfive,m_j=-3/2}\rightarrow\ket{\Bs,m_j=-5/2}$ transition at low power for a variable time $t_{434}$. The steady-state population will accumulate in only two states, $\ket{\Dfive,m_j=-5/2}$ and $ \ket{\Dthree,m_j=-3/2}$ as shown in Fig.~\ref{fig:Branching}. The next closest Zeeman transition relative to $\ket{\Dfive,m_j=-3/2}$, corresponding to $\ket{\Dfive,m_j=-3/2}\rightarrow\ket{\Bs,m_j=-3/2}$, is 6 MHz away and thus not driven at sufficiently low power. We then wait for 500 $\mu$s to allow for the decay of the $\Bs$ state to be complete before detecting the probability of populating the $\Shalf$ and $\Dthree$ states (see Supplemental Material). 
If we assume that the decays to the $\Shalf$ and $\Fseven$ states are negligible, the steady-state population can be approximated as:
\begin{eqnarray}
\label{eq:branching}
P_{ss}( ^2D_{3/2})&=&\frac{b_{D_{3/2}}}{b_{D_{3/2}}+b_{D_{5/2}\ket{m=-5/2}}},\\
b_{D_{5/2}\ket{m=-5/2}}&=&|\bra{5/2,-5/2;1,0}5/2,-5/2\rangle|^2 b_{D_{5/2}}.\nonumber
\end{eqnarray}
Here, $b_{D_{5/2},\ket{m_j=-5/2}}$ is the branching ratio of $\ket{\Bs,m_j=-5/2} \rightarrow \ket{D_{5/2},m_j=-5/2} $ weighted by the relevant Clebsch-Gordan coefficient $\bra{5/2,-5/2;1,0}5/2,-5/2\rangle$. However, there is a slight increase of the measured population at long times due to slow decays of the $\Dfive$ ($\tau_{D_{5/2}}=7.2(3) $ ms \cite{Taylor1997}) and $\Dthree$ states ($\tau_{D_{3/2}}=54.83(18)$ ms \cite{Shao2023}) to the $\Shalf$ and $\Fseven$ states (Fig.~\ref{fig:Branching}b) that is not captured by Eq.~\eqref{eq:branching}. To take these corrections into account, we perform a numerical simulation of the OBEs (see Supplemental Material), including all the atomic levels involved as well as the known decay rates of $\Dthree$ and $\Dfive$ states. We perform the measurements at four different laser powers with Rabi frequencies (see inset in Fig: \ref{fig:Branching}b) much smaller than the Zeeman splitting to the closest transition but larger than $\gamma_{[5/2]}/2\pi$. All four datasets converge to the same long-time behaviour, confirming the low-power regime. Using the measured lifetime as a fixed parameter, the total branching ratio to $\Dthree$ and $\Shalf$ states given by the fit to the full OBEs is $b_{D_{3/2}}+b_{S_{1/2}}=0.361(2)$.

In order to quantify the direct branching ratios to the $\Shalf$ and $\Fseven$ states ($b_{S_{1/2}}$ and $b_{F_{7/2}}$ in the following), we drive the $\Dthree\rightarrow\Bs$ and $\Dfive\rightarrow\Bs$ transitions at high power ($\Omega_{410}/2\pi \approx 3$ MHz and $\Omega_{434}/2\pi \approx 7.8$ MHz) and low magnetic field ($B= 0.448(2)$ Gauss), such that the Rabi frequencies are much larger than the total Zeeman splitting bandwidths, where $\Delta f^{(410)}_{z}=1.636$ MHz and $\Delta f^{(434)}_{z}=1.974$ MHz separate the two extreme Zeeman transitions, respectively. In this regime, all the Zeeman states of the $\Dthree$($\Dfive$) manifold are depopulated during the 410(434) nm pulse. By measuring the total population in $\Shalf+\Dthree$ as a function of 410(434) nm pulse time, we can fit the data assuming the previously estimated values of $\gamma_{[5/2]}$ and $b_{D_{3/2}}+b_{S_{1/2}}$ and leaving $b_{S_{1/2}}$ ($b_{F_{7/2}}$) as the only fit parameter.


In the case of $b_{S_{1/2}}$, we use the 410 nm high-power pumping scheme (Fig.~\ref{fig:Branching}d). As shown in Fig.~\ref{fig:Branching}e, any population in addition to the direct $\Dthree,\Dfive\rightarrow \Shalf$ decays can be attributed to the direct E2 $\Bs\rightarrow\Shalf$ decay. The comparison of the data (purple points) with the numerical results assuming $b_{S_{1/2}}=0$ (dashed line) suggests a non-zero value for the branching ratio of this E2 decay. Since the uncertainty of $b_{S_{1/2}}$ is limited by the known accuracy of the lifetime and branching ratio of the $\Dfive\rightarrow\Shalf$ decay, we performed a separate measurement of the lifetime $\tau_{D_{5/2}}=7.3(3)$ and branching ratio $b_{D_{5/2}\rightarrow S_{1/2}}=0.188(3)$. While our measurement of $\tau_{D_{5/2}}$ is in agreement with previous works on \Ybc ions \cite{Taylor1997}, we improved the relative uncertainty of $b_{D_{5/2}\rightarrow S_{1/2}}$ by an order of magnitude. Combining this uncertainty with other systematic effects (see Appendix), we obtain $b_{S_{1/2}}=0.0023(16)$.

On the other hand, the blue data points, measured using the 434 nm pumping scheme shown in Fig.~\ref{fig:Branching}c, closely follow the simulation assuming $b_{F_{7/2}}=0$. The fit result of $b_{F_{7/2}}=0.0000(2)$ is also statistically compatible with zero, inferring that the missing population in the steady state can be entirely attributed to the direct $\Dfive\rightarrow\Fseven$ decay. Considering the error budget analysis (see Appendix), our experimental resolution is insufficient for precisely determining the branching ratio to $\Fseven$. 

For both pumping schemes, we independently determine the Rabi frequencies by fitting coherent oscillations at short times (see Supplemental Material). The estimated values of $b_{F_{7/2}}$ and $b_{S_{1/2}}$ yield $b_{D_{3/2}}=0.359(2)$ and $b_{D_{5/2}}=0.639(2)$.

\emph{$\Bs$ state Landé g-factor measurement} - To estimate the Landé g-factor, we measure the frequencies of several Zeeman transitions of $\Shalf \rightarrow \Dfive$ and $\Dfive \rightarrow \Bs$ using low-power Rabi spectroscopy at 411 nm and 434 nm at different magnetic fields (see Appendix). 
Once the ratio of Zeeman splittings is known, we can estimate the Landé g-factor of the $\Bs$ state as follows:
 \begin{equation}
     \frac{g_{D_{5/2}}}{g_{S_{1/2}}}=\frac{\Delta f_{D_{5/2}}}{\Delta f_{S_{1/2}}}, \,\,
      \frac{g_{[5/2]_{5/2}}}{g_{S_{1/2}}}=\frac{\Delta f_{[5/2]_{5/2}}}{\Delta f_{S_{1/2}}},
 \end{equation}
where $\Delta f_{a}$ is the Zeeman splitting between states with $\Delta m_{j}=1$ for $a=\{\Shalf,\Dfive, \Bs\}$. The ratio between the Zeeman splittings of $\Dfive$ and $\Shalf$ ($\Bs$ and $\Shalf$) is $0.5994(2)$ ($0.4922(2)$), where the uncertainty is mainly due to magnetic field fluctuations (see Appendix).
This ratio can be used along with previous $\Shalf$ or $\Dfive$ measurements \cite{Meggers1967} or theoretical predictions \cite{Han2025} to estimate the Landé g-factor of $\Bs$. Assuming $g_{S_{1/2}}=1.998$ \cite{Meggers1967}, we obtain $g_{D_{5/2}}=1.1976(4)$ and $g_{[5/2]_{5/2}}=0.9834(4)$. Alternatively, assuming the recent theoretical estimate \cite{Han2025} of $g_{S_{1/2}}=2.002615$, we obtain $g_{D_{5/2}}=1.2003(4)$ and $g_{[5/2]_{5/2}}=0.9856(4)$.

\emph{Discussion} - In this work, we have reported the first precision measurement of the lifetime, branching ratios, and g-factor of the high-lying $\Bs$ state in \Ybc ions. The relatively long lifetime measured here questions the degree of mixing between the $4f^{13}6s5d$ and $4f^{13}5d^2$ configurations in the $\Bs$ state \cite{NISTdatabase}. Our work also paves the way to new schemes for \Yb qubit manipulation in the $omg$ architecture, as it shows how high-lying ``bracket'' states can be used for quantum information processing. The fact that the $\Dfive$,$\Dthree$, and $\Bs$ states form a quasi-closed lambda system (up to the slow decays to the $\Shalf$ and $\Fseven$ states) suggests that high-lying states might be used for mid-circuit measurements by scattering photons directly in the metastable $4f^{14}5d$ manifold. For example, \Yb ions feature multiple odd parity states belonging to the $4f^{13}5d^2$ configuration that might exhibit broader linewidths and closed cycling transitions, more suitable for photon collection and high-fidelity detection than the $\Bs$ state studied here.
Through the $\Bs$ state, one can also pump the $\Dthree$ population into the $\Dfive$ state, circumventing the fundamental limit to the shelving fidelity represented by the M1 decay from the $\Dfive$ to the $\Dthree$ state during optical pumping with 411 nm light \cite{Conrad2021, Edmunds2021}. Additionally, the presence of potentially three excitation paths, 
all leading to the relatively narrow-linewidth $\Bs$ state, opens intriguing possibilities in tests of the foundation of quantum mechanics \cite{Weinberg2016, Raizen2022}.

Finally, we point out that there are currently no numerical methods available to predict and explain our observations. Our work will therefore serve as a testbed to benchmark numerical methods and to better understand the atomic structure of \Yb ions, a workhorse of quantum information science.

%

\emph{Acknowledgments}:
The authors are grateful to Marianna Safronova, Sergey Porsev, and Julian Berengut for insightful discussions and to Luis Orozco for critical reading of the manuscript. G.P. acknowledges the support of the Office of Naval Research (Grants No. N00014-23-1-2665 and N00014-24-12593), the Welch Foundation Award C-2154, the Office of Naval Research Young Investigator Program (Grant No. N00014-22-1-2282), and the NSF CAREER Award (Grant No. PHY-2144910). We acknowledge that this material is based on work supported by the U.S Department of Energy, Office of Science, Office of Nuclear Physics under the Early Career Award No. DE-SC0023806.

\bibliographystyle{bracket_sty}
\bibliography{main}
\section{Appendix}

\subsection{Systematic Errors}

The systematic errors for the quantities measured in this work are listed in Table \ref{table_error}.
All measurements are affected by frequency fluctuations of the relevant atomic transitions, induced by magnetic field variations, drifts in the ULE cavity, and laser intensity variations. To quantify the magnetic field fluctuations, we measure a magnetically sensitive microwave transition in the ground state of a \Ybf ion over the span of 30 minutes, observing a slow oscillatory fluctuation with a fractional rms instability of $0.5 \cdot 10^{-5}$. We also characterize the ULE drift by measuring the two clock transitions of $\Shalf\rightarrow\Dthree,\Dfive$ over several months and find it to be 0.175 Hz/s. The Stark shift of the 410 nm transition could not be resolved in the power range explored in this work, in agreement with the OBE model predicting a few 10s of kHz at maximum power. In the case of the 434 nm transition, we verified the OBE model prediction experimentally ($\sim 150$ kHz at the largest power) and characterized the fluctuations of the laser power to be 2\% over the course of a typical experimental run. Another source of systematic error is the rms fluctuations of the SPAM fidelity over time, which we measured to be $1\cdot10^{-3}$.

Furthermore, we fit the data to an OBE model, which uses the lifetimes of the $\Dthree$ and $\Dfive$  states ($\tau_{D_{3/2}}$ and $\tau_{D_{5/2}}$) and the branching ratio of the $\Dfive\rightarrow\Shalf$ decay ($b_{D_{5/2}\rightarrow S_{1/2}}$) as inputs. Therefore, we also have to consider their known uncertainties~\cite{Taylor1997, Shao2023}. For $b_{D_{5/2}\rightarrow S_{1/2}}$, we use the value and uncertainty measured in this work.

Finally, we consider the truncation errors caused by fitting only selected portions of the datasets. We neglect the systematic effects of quantum beats, as our measurements are largely insensitive to them because we detect the population resulting from the decay tracing over the final Zeeman states and not the photons emitted by the decay. 

\begin{table}[t!]
\centering
\caption{{\bf Error budget for lifetime $\tau$ and branching ratios $b_{D_{3/2}}$, $b_{S_{1/2}}$, and $b_{F_{7/2}}$}. The errors on the branching ratios are in absolute percentages. }
\begin{tabular}{*6c}
 &  \multicolumn{1}{c}{$\tau$}\, & \multicolumn{1}{c}{$b_{D_{3/2}}$} &  \multicolumn{1}{c}{$b_{F_{7/2}}$} & \multicolumn{1}{c}{$b_{S_{1/2}}$} \\
\hline
 Noise source         & ms  & \%  & \% & \%  \\
\hline
\hline
$\tau$                    & --  & 1.1e-2  &  6.9e-3 &1.4e-3  \\
ULE cavity drift          & 2.1e-4  & 3.6e-3  & 1.6e-7  &  5.8e-7   \\
Magnetic field            & 1.3e-8  & 1.6e-3   & 8.4e-3  &3.2e-3 \\
Stark Shift                         & 1.9e-4  &  2.1e-3   & 4.8e-7 &  8.5e-5 \\
SPAM                                & 1.3e-8  & 4.0e-2  & 3.7e-3 &4.4e-2  \\
$\tau_{D_{3/2}}$, $\tau_{D_{5/2}}$, $b_{D_{5/2}\rightarrow S_{1/2}}$ & 2.8e-4  & 0.14  &   6.9e-3  & 0.15\\     
Truncation errors                   &  3.2e-4 & 1.5e-2  &1.2e-2  & 1.4e-2 \\
\hline
Total Systematic & 5.2e-4  & 0.15&  1.8e-2 & 0.16 \\
\hline
Statistical & 7.8e-4  & 8.8e-2&  2.0e-2 & 2.9e-2 \\
\hline
\hline
Total & 0.0009  & 0.17 &2.6e-2 &0.16 \\
\hline
\hline
\vspace{-0.5cm}
\label{table_error}
\end{tabular}\end{table}

To estimate the systematic errors for the reported quantities, we use the sources of error
(ULE frequency, magnetic field, Stark shift, SPAM, and $\tau_{D_{3/2}}$, $\tau_{D_{5/2}}$, $b_{D_{5/2}\rightarrow S_{1/2}}$) as Gaussian random variables with experimentally characterized variance to perform Monte Carlo sampling and fit the measured data with each Monte Carlo sample. With the resulting samples of fitted quantities $\{\tau, b_{D_{3/2}}, b_{D_{5/2}},b_{S_{1/2}},b_{F_{7/2}}\}$, we extract the weighted standard deviation to estimate the systematic error of that particular parameter. 
Similarly, when appropriate, we perform a truncation analysis to extract the systematic effect on how we select the data subset. In the case of the lifetime measurement (Fig.~\ref{fig:lifetime}), we consider 13 truncation windows from $\sim 3\tau$ to the last timestamp, increasing by $10.5\,\mu$s per window. For the measured branching ratio $b_{D_{3/2}}+b_{S_{1/2}}$ (Fig.~\ref{fig:Branching}b), we truncate from $320\,\mu$s to the last timestamp, using 9 truncation windows, each with a span of $80\,\mu$s. 
We also truncate the initial point from 0.5 ms to 1.5 ms with 20 truncation windows ($5\,\mu$s each) and from 0.5 ms to 1.25 ms with 12 truncation windows ($12.5\,\mu$s each) for $b_{F_{7/2}}$ and $b_{S_{1/2}}$, respectively.

Finally, we note that we verify with Monte Carlo sampling that the experimental protocol of the lifetime measurement is insensitive to the values of the branching ratios($<1.1\cdot 10^{-4}$, not reported in Table \ref{table_error}), as their variations would affect the scale of the overall signal and not the decay time. At the same time, in the case of the datasets used to extract the branching ratios (Fig.~\ref{fig:Branching}b), we truncate the data to make sure that the first part of the decay, which is more sensitive to the value of $\tau$, is removed, and we measure the branching ratios using only the plateau at the end of the evolution that is mostly affected by the branching ratios and the values of $\tau_{D_{3/2}}$, $\tau_{D_{5/2}}$, and $b_{D_{5/2}\rightarrow S_{1/2}}$. We check that this procedure guarantees a systematic error induced by the uncertainty of $\tau$ for the branching ratios of $<1.1\cdot 10^{-4}$ in absolute scale, as shown in Table \ref{table_error}.

\subsection{Landé g-Factor Measurement}
\begin{figure}[t!]
    \centering
    \includegraphics[width=0.5\textwidth]{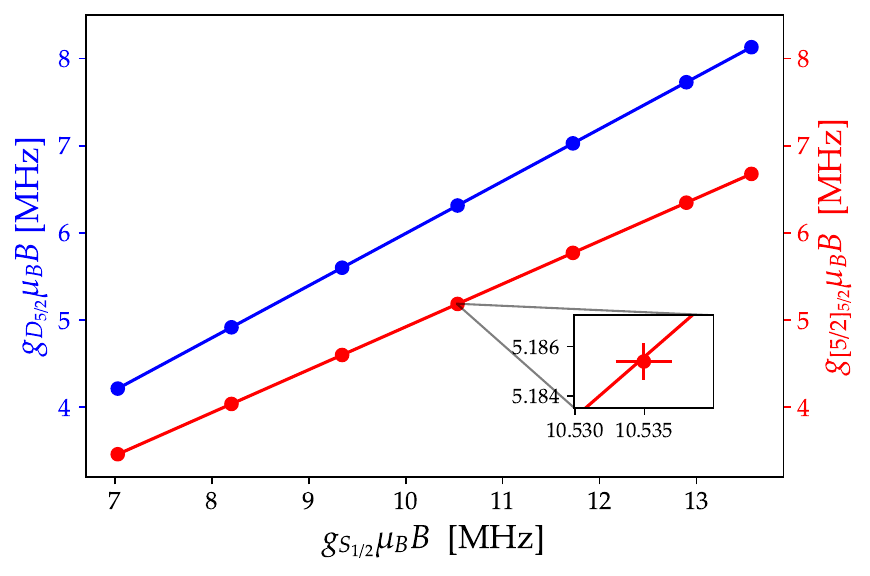}
    \caption{Relative Zeeman frequency splitting of $\Dfive$ and $\Bs$ as a function of $\Shalf$ Zeeman splitting for different magnetic fields. The solid line is a linear fit of the experimental values. The error bars are shown in the inset.}
    \vspace{-0.5cm}
    \label{fig:landeg}
\end{figure}

The magnetic field is controlled by tuning the current in the electromagnetic coils placed symmetrically around the vacuum chamber in the Helmholtz configuration.
For each magnetic field, we first use the 411 nm beam to measure the transition frequencies from both the $\ket{\Shalf,m_j=-1/2}$ and $\ket{\Shalf,m_j=1/2}$ to the $\ket{\Dfive,m_j'=-3/2,-1/2,1/2}$ and $\ket{\Dfive,m_j'=-1/2,1/2,3/2}$ Zeeman states, respectively. Each measurement is quickly followed by a similar spectroscopy at low 434 nm power, starting from $\ket{\Dfive,m_j=-1/2}$ and $\ket{\Dfive, m_j=1/2}$ to $\ket{\Bs,m_j'=-3/2,-1/2,1/2}$ and $\ket{\Bs, m_j'=-1/2,1/2,3/2}$ states, respectively. Finally, the $\Shalf \rightarrow \Dfive$ transition is measured again to account for drifts in the magnetic field and ULE cavity, and the average of the two values is taken for each subtransition.  The observed Zeeman splittings are shown in Fig.~\ref{fig:landeg}.

\subsection{Measurement of Lifetime and Branching ratio of the $\Dfive$ state}
To measure the lifetime of $\Dfive$ and its branching ratio to $\Shalf$, we prepare the ion in the $\Dfive |m_j=-3/2\rangle$ state using the heralding procedure and observe the total population in $\Shalf$ and $\Dthree$ as a function of wait time. We measure the $\Dfive$ lifetime of 7.3(3) ms and the branching ratio from $\Dfive$ to $\Shalf$ to be 0.188(3). The error bars include the systematic effects of SPAM fluctuations.
\begin{figure}[t!]
    \centering
    \includegraphics[width=0.5\textwidth]{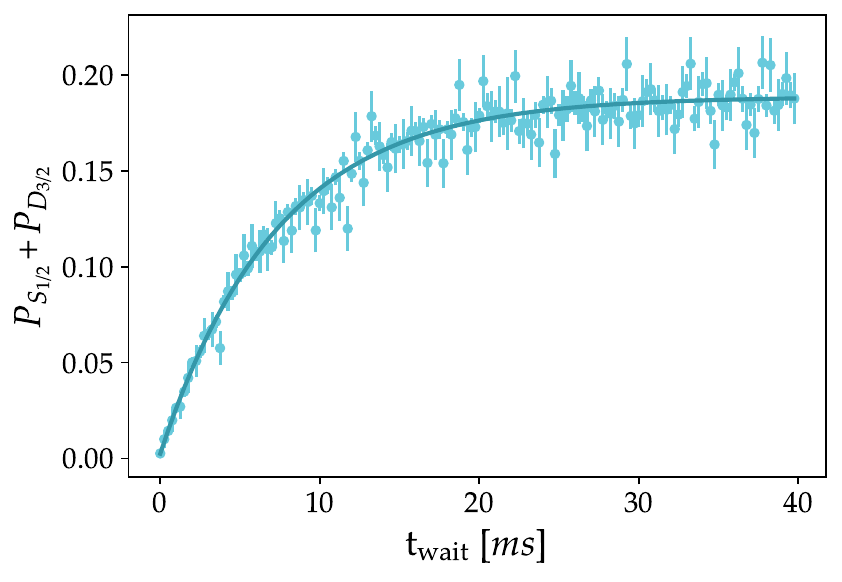}
    \caption{Decay of Population from $\Dfive$ to $\Shalf$. Each point is the average of 4000 repetitions taken over multiple datasets. The solid line is the fit to an exponential curve. }
    \vspace{-0.5cm}
    \label{fig:D52S12decay}
\end{figure}

\section{Supplemental Material}

\subsection{Master Equation}
The Optical Bloch Equations (OBE), describing the system including the Zeeman levels of $\Bs, \Dfive$, and $\Dthree$, are numerically integrated using QuTiP \cite{johansson2013qutip}. They can be written as a Lindblad master equation: 
\begin{eqnarray}
    \frac{d\hat{\rho}}{dt}=-\frac{i}{\hbar}[\hat{H},\hat{\rho}]+\mathcal{L}(\hat{\rho}),
\end{eqnarray}
where $\hat{H}$ is the Hamiltonian operator describing the unitary part of the evolution, including Rabi frequencies $\Omega_{ij}$ coupling states $\ket{i}$ and $\ket{j}$ and detunings $\Delta_i=\omega_L - \omega_i$, where $\omega_L$ is the laser frequency and $\omega_i$ is the frequency of state $\ket{i}$ within the rotating wave approximation. The Lindbladian superoperator $\mathcal{L}(\hat{\rho})$ takes into account the non-unitary part of the evolution and is given by:
\begin{eqnarray}
    \mathcal{L}(\hat{\rho})=-\frac{1}{2}\sum_k (\hat{C_k^\dagger}\hat{C_k}\hat{\rho}+\hat{\rho}\hat{C_k^\dagger}\hat{C_k}-2\hat{C_k}\hat{\rho}\hat{C_k^\dagger}),
\end{eqnarray}
where $\hat{C_k}$ are the collapse operators used to model the decay from state $\ket{i}$ to state $\ket{j}$:
\begin{eqnarray}
    \hat{C_k}=\sqrt{\Gamma_{ij}}\ket{j}\bra{i}.
\end{eqnarray}
The lifetimes and branching ratios are included in the weights $\Gamma_{ij}$ of the different collapse operators $\hat{C}_k$.

\subsection{Pulse Sequences and Fitting Procedures for Rabi Oscillations and Branching Ratio Measurements}
In this section, we provide details on the pulse sequences used to observe Rabi oscillations and to measure the branching ratios, as explained in Fig.~\ref{fig:Branching}.

\begin{figure}[t!]
    \centering\includegraphics[width=0.5\textwidth]{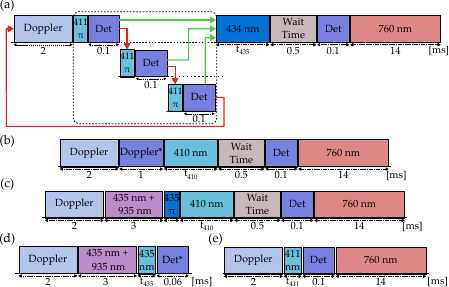}
    \vspace{-0.5cm}
    \caption{{\bf (a)} Pulse sequence for $b_{D_{3/2}}+b_{S_{1/2}}$ and $b_{F_{7/2}}$ measurements with variable 434 nm time (Figs.~\ref{fig:Yblevel}, \ref{fig:Branching}b, \ref{fig:Branching}e (blue points)). {\bf (b)} Pulse sequence for $b_{S_{1/2}}$ measurement with variable 410 nm time (Fig.~\ref{fig:Branching}e, purple points). Here, Doppler* refers to the illumination of 370 nm light without 935 nm light to pump the population into $\Dthree$ state. {\bf (c)} Pulse sequence for $\Dthree\rightarrow \Bs$ Rabi oscillations at 410 nm. {\bf (d,e)} Pulse sequence for $\Shalf\rightarrow (\Dthree,\Dfive)$ Rabi oscillations at (435 nm, 411 nm) (Fig.~\ref{fig:Yblevel}). Here, Det* refers to the illumination of 370 nm light without 935 nm light for state detection.}
    \vspace{-0.5cm}
    \label{fig:pulse_434_410_supp}
\end{figure}

\begin{figure}[t!]
    \centering
\includegraphics[width=0.5\textwidth]{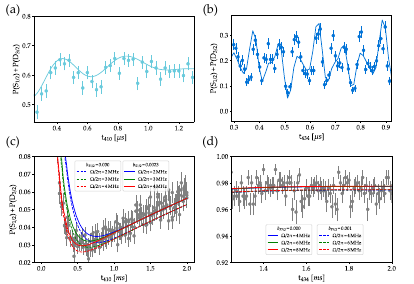}
    \vspace{-0.5cm}
    \caption{{\bf (a,b)} Initial oscillations of 410 (434) nm for $b_{S_{1/2}}$ ($b_{F_{7/2}}$) measurement of Fig.~\ref{fig:Branching}. In the case of 434 nm (b), the beatings are due to multiple Zeeman sublevels participating in the coherent dynamics at high power. The solid line is fit to numerical simulation with decoherence to obtain $\Omega_{410}/2\pi=$ 7790 kHz ($\Omega_{434}/2\pi=$ 2771 kHz). {\bf (c,d)} Numerical simulations for different $\Omega_{410}$ and $\Omega_{434}$, compared with the experimental data.}
    \vspace{-0.6cm}
    \label{fig:omegafit_434_410_supp}
\end{figure}

{\bf  434 nm laser pulse sequences}: 
The pulse sequences to observe $\Dfive\leftrightarrow\Bs$ Rabi oscillations (Fig.~\ref{fig:Yblevel}), to measure $b_{D_{3/2}}+b_{S_{1/2}}$ (Fig.~\ref{fig:Branching}b), and to estimate $b_{F_{7/2}}$ (Fig.~\ref{fig:Branching}e, blue data) are similar, with the 434 nm power and magnetic field being the only differences among them (see Fig.~\ref{fig:pulse_434_410_supp}). We prepare $\ket{\Dfive,m_j=-3/2}$ using the heralding protocol, which consists of 411 nm pi pulses followed by detection, repeated up to three times. If successful, we drive the $\ket{\Dfive,m_j=-3/2}\rightarrow\ket{\Bs,m_j=-5/2}$ transition for a variable time $t_{434}$. Afterwards, we wait for 500 $\mu$s to allow for the decay of the $\Bs$ state to be complete. Finally, we detect the population probability of $\Shalf$ and $\Dthree$ states with 370 nm and 935 nm lasers.
When we drive the $\Dfive\rightarrow\Bs$ transition at high power to estimate $b_{F_{7/2}}$ (Fig.~\ref{fig:Branching}e, blue data), we calibrate the Rabi frequency $\Omega_{434}$ using the data at short times (see Fig.~\ref{fig:omegafit_434_410_supp}b). We emphasize that, since the $\Dfive\rightarrow\Bs$ transition is strongly saturated, and $\Omega_{434}$ is larger than the Zeeman bandwidth, the estimate of $b_{F_{7/2}}$ depends mostly on the steady-state value of the population probability. Therefore, it is largely insensitive to the specific value of $\Omega_{434}$ (as shown in Fig.~\ref{fig:omegafit_434_410_supp}d) and to the short-time dynamics.

{\bf 410 nm laser pulse sequences: }
To observe 410 nm Rabi oscillations, we first drive the $\Delta m_j=+1$ transition of $\ket{\Shalf, m_j=-1/2}\rightarrow \ket{\Dthree,m_j=+1/2}$ at 435 nm together with the 935 nm repumper for 2 ms. This results in optical pumping in the $\ket{\Shalf,m_j=+1/2}$ state.
We then prepare the $\ket{\Dthree,m_j=+3/2}$ state with a 435 nm $\pi$ pulse and drive the $\ket{\Dthree,m_j=+3/2}\rightarrow\ket{\Bs,m_j=+5/2}$ transition. After waiting 500 $\mu$s to allow for the decay of the $\Bs$ state to be complete, we detect the population probability of $\Shalf$ and $\Dthree$ states with 370 nm and 935 nm lasers. 

In the case of optical pumping evolution caused by the strong drive of the $\Dthree\rightarrow\Bs$ transition used to estimate $b_{S_{1/2}}$ (Fig.~\ref{fig:Branching}e, purple data), it is crucial to prepare the population in the $\Dthree$ state with high fidelity. However, since the detection step mixes the $\Dthree$ and $\Shalf$ states, the heralded state preparation is not possible with a 435 nm excitation. Therefore, to maximize the state-preparation fidelity here, we prepare a mixture of all $\Dthree$ Zeeman sublevels by applying a Doppler pulse without 935 nm. Since the Rabi frequency is larger than the Zeeman bandwidth at low magnetic field ($B=0.448(2)$ Gauss), a specific pure-state initialization in $\Dthree$ is not necessary, as we can set the 410 nm laser at the peak of the broadened Zeeman spectrum.

Similar to the case of the high power drive at 434 nm, the pumping characteristic timescale with a strong 410 nm drive is much shorter than the slow E2 $\Dfive,\Dthree\rightarrow\Shalf$ decays ($\tau_{D_{5/2}}=7.2(3) $ ms \cite{Taylor1997} and $\tau_{D_{3/2}}=54.83(18)$ ms \cite{Shao2023}). In this high-power regime, the slow, linear increase at long times ($t>0.6$ ms) is mostly determined by $\tau_{D_{5/2}}$, while the offset of the linear slope is determined by $b_{S_{1/2}}$. Therefore, we choose to fit only the points at $t>0.6$ ms to minimize the dependence of our estimate on the precise value of the Rabi frequency $\Omega_{410}$, which is independently calibrated in Fig.~\ref{fig:omegafit_434_410_supp}a. In Fig.~\ref{fig:omegafit_434_410_supp}c, we show the result of multiple numerical simulations at Rabi frequencies in the range of $\Omega_{410}/2\pi=2-4$ MHz, assuming either $b_{S_{1/2}}=0$ or $b_{S_{1/2}}=0.0023$, showing how the long time dynamics is insensitive to the specific value of the Rabi frequency and that the branching ratio $b_{S_{1/2}}$ determines the offset common to all the curves.

{\bf 435 nm and 411 nm Rabi oscillations:}
To observe Rabi oscillations on the two E2 transitions ($\Shalf\rightarrow\Dthree,\Dfive$) in Fig.~\ref{fig:Yblevel}, we prepare an initial state in the Zeeman $\Shalf$ manifold with the 435 nm + 935 nm pumping protocol and then illuminate the ion with 435 (411) nm laser, resonant to the $\ket{\Shalf, m_j=+1/2}\rightarrow\ket{\Dthree, m_j=+1/2}(\ket{\Dfive,m_j=+1/2})$ transition, for a variable time. For $\Shalf\rightarrow\Dthree$, before detection, we use the 411 nm laser to shelve the population from $\Shalf$ to $\Dfive$ state, addressing multiple Zeeman transitions to maximize the shelving fidelity. The population in $\Dthree$ is then detected using both 370 nm and 935 nm lasers. In Fig.~\ref{fig:Yblevel}, no data points are shown between $t=0$ and $t=250$ ns because our control system does not support pulses shorter than 250 ns.

\end{document}